\documentclass[final,5p,times,twocolumn,sort&compress]{elsarticle}
\usepackage{amsmath}
\usepackage{amssymb}
\usepackage{natbib}
\usepackage{graphicx}
\usepackage{physics}
\usepackage{url}
\usepackage{color}
\usepackage{xcolor}
\usepackage{appendix}

\def\bk{{\bf k}}
\def\bR{{\bf R}}
\def\beq{\nopagebreak \begin{equation}}
\def\eeq{\end{equation}}
\def\opbraket#1#2#3{\langle #1|#2|#3\rangle}

\def\half{\frac{1}{2}}
\def\id{{\textrm d}}

\graphicspath{{images/}}

\newcounter{bla}

\journal{Computer Physics Communications}
\begin{document}
\begin{frontmatter}

\title{\texttt{BoltzTraP2}, a program for interpolating band structures and calculating semi-classical transport coefficients}
\author[tuw]{Georg K. H. Madsen\corref{cor1}}
\author[tuw]{Jes\'us Carrete}
\author[uli]{Matthieu J. Verstraete}

\cortext[cor1]{Corresponding author.\\\textit{E-mail address:} georg.madsen@tuwien.ac.at}
\address[tuw]{Institute of Materials Chemistry, TU Wien, A-1060 Vienna, Austria}
\address[uli]{nanomat/QMAT/CESAM and Department of Physics, Universit\'e de Li\`ege, all\'ee du 6 ao\^ut, 19, B-4000 Li\`ege, Belgium.}

\begin{abstract}
\texttt{BoltzTraP2} is a software package for calculating a smoothed Fourier expression of periodic functions and the Onsager transport coefficients for extended systems using the linearized Boltzmann transport equation. It uses only the band and $k$-dependent quasi-particle energies, as well as the intra-band optical matrix elements and scattering rates, as input. The code can be used via a command-line interface and/or as a Python module. It is tested and illustrated on a simple parabolic band example as well as silicon. The positive Seebeck coefficient of lithium is reproduced in an example of going beyond the constant relaxation time approximation.
\end{abstract}

\end{frontmatter}

{\bf PROGRAM SUMMARY}

\begin{small}
  \noindent
      {\em Program Title:} BoltzTraP2\\
      {\em Journal Reference:}\\
      {\em Catalogue identifier:}\\
      {\em Licensing provisions:} GPLv3+\\
      {\em Programming language:} Python and C++11\\
      {\em Computer:} non-specific\\
      {\em Operating system:} Linux, macOS\\
      {\em RAM:} up to ten GB\\
      {\em Number of processors used:} Typically from 1 to 4
      {\em Classification:} 7.9 Transport Properties\\
      {\em External routines/libraries:} NumPy, SciPy, Matplotlib, spglib, ase, fftw, VTK\\
      {\em Nature of problem:} Calculating the transport coefficients using the linearized Boltzmann transport equation within the relaxation time approximation.\\
      {\em Solution method:} Smoothed Fourier interpolation\\
      {\em Running time:} Up to one hour\\
\end{small}

\section{Introduction}
The first \texttt{BoltzTraP} program \cite{BoltzTraP} provided a numerically stable and efficient method for obtaining analytic representations of quasi-particle energies. It has found broad adoption for the application of the Boltzmann transport equation (BTE) to such diverse fields as superconductors \cite{Singh_PRL08}, transparent conductors \cite{Hautier_NatComm13} inter-metallic phases \cite{Dolinsek_PRB09} as well as thermoelectrics. Its application has been especially widespread for thermoelectrics research \cite{May_PRB09,Ouardi_PRB10,Parker_PRL13,Luo_JMCA14,Kim_NL15,Hong_SciRep16,He_PRL16,Zhang_NatComm16} for which it was originally conceived \cite{Madsen_PRB03,Madsen_JACS06}. Furthermore, it has served as a reference for other methods obtaining transport coefficients, such as maximally localized Wannier functions.\cite{Yates_PRB07,Pizzi_CPC14} The numerical stability means that the approach can be automated \cite{Madsen_JACS06} and the methodology has subsequently been used in several high-throughput studies \cite{Madsen_JACS06,Hautier_NatComm13,Carrete_AFM14,Bhattacharya_PRB15,Zhu_JMCC15}.

The usefulness of the \texttt{BoltzTraP} approach can partly be attributed to the numerical efficiency of the procedure when the quasi-particle energies are approximated by the Kohn-Sham (KS) eigenvalues \cite{KS}. Once the multiplicative potential has been self-consistently calculated, calculating eigenvalues on a fine $\bk$-mesh is a comparatively simple computational step that can be trivially parallelized. An alternative approach calculates the derivatives necessary for the BTE directly from the intra-band momentum matrix elements \cite{Scheidemantel_PRB03}. However, within KS theory, it is often simpler to calculate a finer mesh of eigenvalues than to calculate the momentum matrix elements. When the quasi-particle energies cannot be calculated using a fixed multiplicative KS potential, as in beyond-KS methods such as hybrid functionals \cite{b3} or the GW approach \cite{GW}, this argument no longer holds and calculating the momentum matrix elements \cite{Scheidemantel_PRB03} or using alternative interpolation methods \cite{Prendergast_PRB09,Pizzi_CPC14,Berland_CMS17} could be an advantage.

With the release of \texttt{BoltzTraP2} we wish to achieve three objectives. First of all, a method to use both the eigenvalues and momentum matrix elements is introduced. This ensures that the interpolated manifolds exactly reproduce both the value and derivative at the calculated points. The advantages of the interpolation scheme of the original \texttt{BoltzTraP} approach \cite{BoltzTraP}, and the advantages of using the intra-band momentum matrix elements \cite{Scheidemantel_PRB03} are combined, and the method becomes more suited for beyond-KS approaches. Secondly, we wish to make it more straightforward to avoid the constant relaxation time approximation (RTA) and handle e.g. a temperature-dependent transport distribution function due to electron-phonon coupling \cite{Xu_PRL14,Li_PRB15}. Finally, a further motivation for rewriting and rereleasing \texttt{BoltzTraP} is to provide a modular code based on the modern scripting language Python 3. While \texttt{BoltzTraP} is mainly thought of as a tool to evaluate the transport coefficients, the underlying algorithm can be generally useful for interpolating any periodic function. We hope that the new code can also serve as a library for further developments in this domain.

The paper is built as follows. First we present the interpolation scheme as well as the RTA-BTE. We discuss the interface to the code and provide an outlook and finally we use three examples to illustrate the methodology and results of the code. 

\section{Background}
\subsection{Band interpolation}
The method is based on writing the quasi-particle energies and their derivatives, for each band, as Fourier sums
\begin{gather}
\tilde{\varepsilon}_\bk=\sum_\Lambda c_\Lambda\sum_{R\in\Lambda} \exp(i\bk\cdot\bR)
\label{eq:Fourier} \\
\nabla \tilde{\varepsilon}_\bk=i\sum_\Lambda c_\Lambda\sum_{R\in\Lambda} \bR \exp(i\bk\cdot\bR)
\label{eq:dFourier} 
\end{gather}
where $\Lambda$ are so-called stars representing a set of symmetry-equivalent lattice vectors. 
\texttt{BoltzTraP} was based on the idea by Shankland \cite{Euwema_PR69,Shankland_IJQC71,Koelling_JComP86} that the coefficients should be obtained by minimizing a roughness function under the constraints that calculated quasi-particle energies should be exactly reproduced. This in turn means that the number of coefficients should be larger than the number of calculated points.

The derivatives can also be obtained from the intra-band optical matrix elements \cite{Scheidemantel_PRB03,WOPTIC}
\beq
\nabla \varepsilon_\bk = -\opbraket{\psi_\bk}{\bf p}{\psi_\bk}
\label{eq:mommat}
\eeq
In \texttt{BoltzTraP2} the Shankland algorithm\cite{Euwema_PR69,Shankland_IJQC71,Koelling_JComP86} is extended so that the coefficients ensure that both the quasi-particle energies and their derivatives, Eq.~\eqref{eq:mommat}, are exactly reproduced.
This corresponds to minimizing the Lagrangian
\beq
  I=\half\sum_\bR c_\bR \rho_\bR + 
  \sum_\bk \left[\lambda_\bk \left(\varepsilon_\bk-\tilde{\varepsilon}_\bk\right)
 + \sum_\alpha\lambda_{\alpha,\bk} \nabla_\alpha \left(\varepsilon_\bk- \tilde{\varepsilon}_\bk\right)\right]
\label{eq:Lagrangian}
\eeq
with respect to the Fourier coefficient ($c_{R}$), and choosing the Lagrange multipliers ($\lambda_\bk$ and $\lambda_{\alpha,\bk}$) so that the constraints are fulfilled. The index $\alpha$ labels the three Cartesian directions and indicates that each calculated derivative, Eq.~\eqref{eq:mommat}, adds three Lagrange multipliers.  
Like in the \texttt{BoltzTraP} code, we use the roughness function provided by Pickett et al. \cite{Pickett_PRB88}
\beq
\rho_\bR=\left( 1-c_1 \frac{R}{R_{min}}\right)^2 +c_2\left(\frac{R}{R_{min}}\right)^6.
\label{eq:Pickett}
\eeq

\subsection{Boltzmann transport equation}
\texttt{BoltzTraP2} calculates transport coefficients based on the rigid-band approximation (RBA), which assumes that changing the temperature, or doping a system, does not change the band structure. In the following we will suppress the orbital index. In the RBA the carrier concentration, for a given $T$ and $\mu$, in a semiconductor can be obtained directly from the density of states (DOS)
\beq
n(\varepsilon)= \int  \sum\limits_b\delta(\varepsilon-\varepsilon_{b,\bk}) \frac{\id\bk}{8\pi^3},
\label{eq:dos}
\eeq
where the subscript $b$ runs over bands, by calculating the deviation from charge neutrality
\beq
c(\mu,T)=N-\int n(\varepsilon) f^{(0)}(\varepsilon;\mu,T) \id \varepsilon.
\label{eq:carrier}
\eeq
In Eq.~\eqref{eq:carrier}, $N$ is the nuclear charge and $f^{(0)}$ is the Fermi distribution function. In a semiconductor where charge neutrality would place the Fermi level in the band-gap, one can thus imagine how (at $T=0$) moving $\mu$ into the conduction bands would produce a $n$-type material and moving $\mu$ into the valence bands would produce a $p$-type material.

The BTE describes the behavior of an out-of-equilibrium system in terms of a balance between scattering in and out of each possible state, with scalar scattering rates \cite{Ziman_book}. We have implemented the linearized version of the BTE under the RTA, where the transport distribution function
\beq
\sigma(\varepsilon,T)= \int  \sum\limits_b{\bf v}_{b,\bk} \otimes {\bf v}_{b,\bk} \tau_{b,\bk} \delta(\varepsilon-\varepsilon_{b,\bk}) \frac{\id\bk}{8\pi^3}
\label{eq:sigma}
\eeq
is used to calculate the moments of the generalized transport coefficients
\beq
{\mathcal L}^{(\alpha)}(\mu;T)=q^2\int \sigma(\varepsilon,T) (\varepsilon-\mu)^\alpha\left(-\frac{\partial f^{(0)}(\varepsilon;\mu,T)}{\partial \varepsilon}\right)\id\varepsilon
\label{eq:coeff_E}
\eeq
which give the charge and heat currents
\begin{gather}
j_e=\mathcal{L}^{(0)}{\bf E}+\frac{\mathcal{L}^{(1)}}{qT}(-\nabla T)\\
j_Q=\frac{\mathcal{L}^{(1)}}{q}{\bf E}+\frac{\mathcal{L}^{(2)}}{q^2T}(-\nabla T)
\end{gather}
Identifying the two experimental situations of zero temperature gradient and zero electric current, we obtain the electrical conductivity, the Peltier coefficient, the Seebeck coefficient and the charge carrier contribution to the thermal conductivity as
\begin{gather}
  \sigma=\mathcal{L}^{(0)} \label{eq:cond} \\
  \Pi=\frac{\mathcal{L}^{(1)}}{q\mathcal{L}^{(0)}} \label{eq:Peltier}\\
  S=\frac{1}{qT}\frac{\mathcal{L}^{(1)}}{\mathcal{L}^{(0)}} \label{eq:Seebeck}\\
  \kappa_e=\frac{1}{q^2T}\left[\frac{(\mathcal{L}^{(1)})^2}{\mathcal{L}^{(0)}} -\mathcal{L}^{(2)}\right]. \label{eq:kappae}
\end{gather}
 The main advantage of the \texttt{BoltzTraP} procedure for evaluating the transport coefficients is that it is straightforward to obtain the group velocities from the $\bk$-space derivatives of the quasi-particle energies, Eq.~\eqref{eq:dFourier}.

\texttt{BoltzTraP} is often associated with the constant relaxation time approximation (CRTA). The CRTA means that the Seebeck coefficient and Hall coefficient become independent of the scattering rate \cite{Singh_PRB97}. Therefore, they can be obtained on an absolute scale as a function of doping and temperature in a single scan. The CRTA in combination with the RBA, which makes the group velocities independent of $\mu$ and $T$, also has a computational advantage as it makes the transport distribution function, Eq.~\eqref{eq:sigma} independent of temperature and doping. The temperature and doping dependence of the transport coefficients ${\mathcal L}^{(\alpha)}$, Eq.~\eqref{eq:coeff_E}, is solely due to the Fermi distribution function, and can be obtained via a scan over a fixed transport distribution function.

Clearly the CRTA will have limitations. It only delivers $\sigma$ and $\kappa_e$ dependent on $\tau$ as a parameter. Furthermore, the independence of $S$ and $R_H$ from $\tau$ is known to break down, even qualitatively, for specific cases \cite{Xu_PRL14}. While it is possible to run the original \texttt{BoltzTraP} with a temperature-, momentum- and band- dependent relaxation time, the structure of the code makes it inconvenient, and the functional form is quite limited. \texttt{BoltzTraP2} makes it much more straightforward.

\section{Implementation and interface}


\subsection{General implementation aspects}

\texttt{BoltzTraP2} is implemented in Python 3, using syntax and standard library features that make it incompatible with Python 2. The Fortran code base of the original \texttt{BoltzTraP} was taken as a reference, but the new version was written from scratch. Numerical data is handled internally in the form of arrays, making extensive use of the  \texttt{NumPy} and  \texttt{SciPy} libraries \cite{numpy,scipy}.  \texttt{Matplotlib} \cite{matplotlib} is used for plotting.

Efficiency is an important goal of this new implementation. Despite being implemented in a higher-level language, \texttt{BoltzTraP2} achieves speeds comparable to the original \texttt{BoltzTraP}. There are several factors contributing to this. First, many of the most expensive operations are vectorized to be performed at a lower level by \texttt{NumPy}, avoiding expensive loops. Second, the symmetry-related code, heavy with loops over long lists of lattice points, is written in C++11 and calls routines from the C API of the \texttt{spglib} library \cite{spglib}. The C++11 components are interfaced to Python by a Cython layer \cite{cython}. Third, fast Fourier transforms are delegated to optimized low-level libraries; specifically, the \texttt{pyFFTW} wrapper around \texttt{FFTW} \cite{fftw} is used if available, with the default \texttt{NumPy} wrapper around \texttt{FFTPACK} \cite{fftpack} as a fallback. Finally, certain ``embarrassingly parallel'' loops can be run on several cores thanks to the \texttt{multiprocessing} module in the Python standard library.

\texttt{BoltzTraP2} allows users to save results to files in JSON format, which is both human readable and parseable with a virtually limitless variety of tools and programming libraries. Specifically, there are two different sorts of JSON-formatted files at play. The first kind, \texttt{bt2} files, contain the DFT input, all information about $k$ points, the interpolation coefficients and a dictionary of metadata. The second category, \texttt{btj} files, contain the DFT input, the DOS, the thermoelectric coefficients for all values of temperature and chemical potential, and another metadata dictionary. Those dictionaries comprise several pieces of version information, a creation timestamp, and information about the scattering model used, if applicable. All JSON files are processed using the LZMA-based \texttt{xz} compressor, to drastically reduce the overhead of the text-based format.

The decision to stick with standard formats also affects other outputs of the calculations, stored in plain text with a column layout very similar to the one created by the original \texttt{BoltTraP} \cite{BoltzTraP}. Most existing post-processing and analysis scripts can be adapted with minimal effort.

Regarding input formats, the current version of the code can read the native output of \texttt{Wien2k} \cite{gmcpc, wien2k} and \texttt{VASP} \cite{VASP2}. In the case of a \texttt{VASP} calculation, only the \texttt{vasprun.xml} file is required, while for \texttt{Wien2k} the necessary pieces of information are read from \texttt{case.struct}, \texttt{case.energy} and \texttt{case.scf}. If derivatives of the bands are to be used, the output-file \texttt{case.mommat2} from the \texttt{OPTIC} \cite{OPTIC} program is used. The code is modular enough that support for other DFT codes can be easily implemented. Alternatively, any DFT code can be adapted (or a translation script written) to create output in \texttt{BoltzTraP2}'s own \texttt{GENE} format, designed for maximum simplicity. Examples of files in this format are provided with the source distribution, and it is only slightly modified compared to the original \texttt{BoltTraP} code \texttt{GENE} format.

\texttt{BoltzTraP2} relies on Python's \texttt{setuptools} module for its build system. On many platforms, the program can be installed from source with a \texttt{python setup.py install} command, and no previous manual configuration. Moreover, we have uploaded it to the Python Package Index, so that even the downloading step can be bypassed in favor of a simple \texttt{pip install BoltzTraP2}. A copy of \texttt{spglib} is bundled with the \texttt{BoltzTraP2}, to avoid a separate installation process. Naturally, a C compiler and a C++11-compliant C++ compiler are still needed when building from source.

\subsection{Command-line interface}

\begin{figure}
  \begin{center}
    \includegraphics[width=\columnwidth]{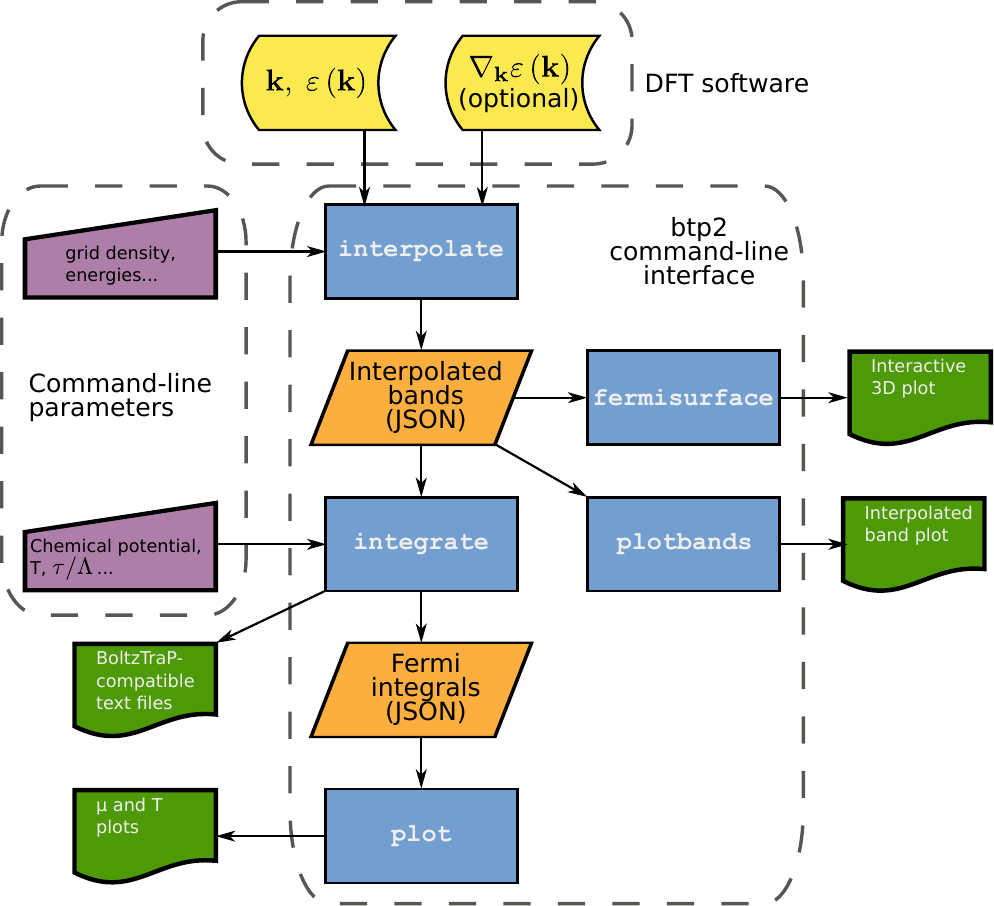}
  \end{center}
  \caption{Typical \texttt{BoltzTraP2} workflow taking the user from the results of a DFT calculation to estimates of the thermoelectric coefficients for the system under study, and other related results, using the \texttt{btp2} command-line interface.}
  \label{fig:workflow}
\end{figure}

The most typical use case of \texttt{BoltzTraP2} is the calculation of transport coefficients. This can be done directly through the \texttt{btp2} command-line front-end, which implements the general workflow depicted in Fig.~\ref{fig:workflow}. It is designed to be self-documenting and controllable through command-line arguments, without the need for configuration files.

The process starts with a set of DFT results, typically from a non-self-consistent calculation using a dense $k$-point grid. The user first calls \texttt{btp2} in ``interpolate'' mode to generate a representation of the bands interpolated to an even denser grid, which is stored in a JSON file. Optional command-line parameters to the ``interpolate'' step can be used to control the grid density, the number of bins used to compute the density of states, the minimum and maximum energies, etc. By saving the result of the interpolation to a file, we avoid repeating the calculation even if the user then wants to generate results for different situations.

To obtain a set of thermoelectric coefficients, the user needs to invoke \texttt{btp2} a second time, now in ``integrate'' mode. In this mode of operation, the command-line script accepts parameters such as the range of temperatures to scan, the dependence of scattering on electron energy, and so on. It then generates output in the form of text files, plus another compressed JSON file as described in the previous section.

The detailed list of command-line parameters, with their meanings, can be obtained by invoking \texttt{btp2} or one of its subcommands with the \texttt{-h} or \texttt{--help} flag.

In addition to the ``integrate'' and ``interpolate'' subcommands, Fig.~\ref{fig:workflow} illustrates the use of the ``fermisurface'', ``plotbands'' and ``plot'' modes of the \texttt{btp2} command-line interface. Their job is to generate graphical representations of the \texttt{BoltzTraP2} output: an interactive 3D plot of the Fermi surface for different values of the energy, a plot of the interpolated electron energies along a specific path in reciprocal space, and plots of the thermoelectric coefficients as functions of temperature and chemical potential, respectively. 3D plotting will only be available if the optional \texttt{vtk} module is detected.

The code is documented, and a set of unit tests covering all the basic functionality is provided with the source distribution. The whole battery of tests can be run with \texttt{pytest}.

\subsection{Using \texttt{BoltzTraP2} as a module}

Advanced users may prefer to skip the command-line interface and access the full feature set of \texttt{BoltzTraP2} more directly. Those wanting to use the interpolation capabilities of \texttt{BoltzTraP2} in their own code, or using it as part of an automated workflow, will probably fall in this category. Furthermore, the \texttt{btp2} command-line interface only allows choosing between a uniform-relaxation-time model and a uniform-mean-free-path one. Users requiring custom parameterizations of the electronic scattering rates will need to bypass the interface. This is easily accomplished by calling the API of the \texttt{BoltzTraP2} Python module, either from a script or from an interactive Python shell, such as the \texttt{Jupyter} notebook \cite{ipython}. Crystal structures are represented as \texttt{ase} atoms objects \cite{ase}, which allows for easy interfacing with many other Python libraries and external software.

The best reference about the API of the \texttt{BoltzTraP2} is the source code of the \texttt{btp2} interface itself, and a set of documented examples that are provided with the source distribution of \texttt{BoltzTraP2}. The examples illustrate how to accomplish specific tasks and reproduce several results obtained with the original \texttt{BoltzTraP} code as well as the three examples in the following section.

\section{Examples}
\subsection{Isotropic parabolic band model}
\begin{figure}[htb]
\begin{center}
\includegraphics[width=.9\columnwidth]{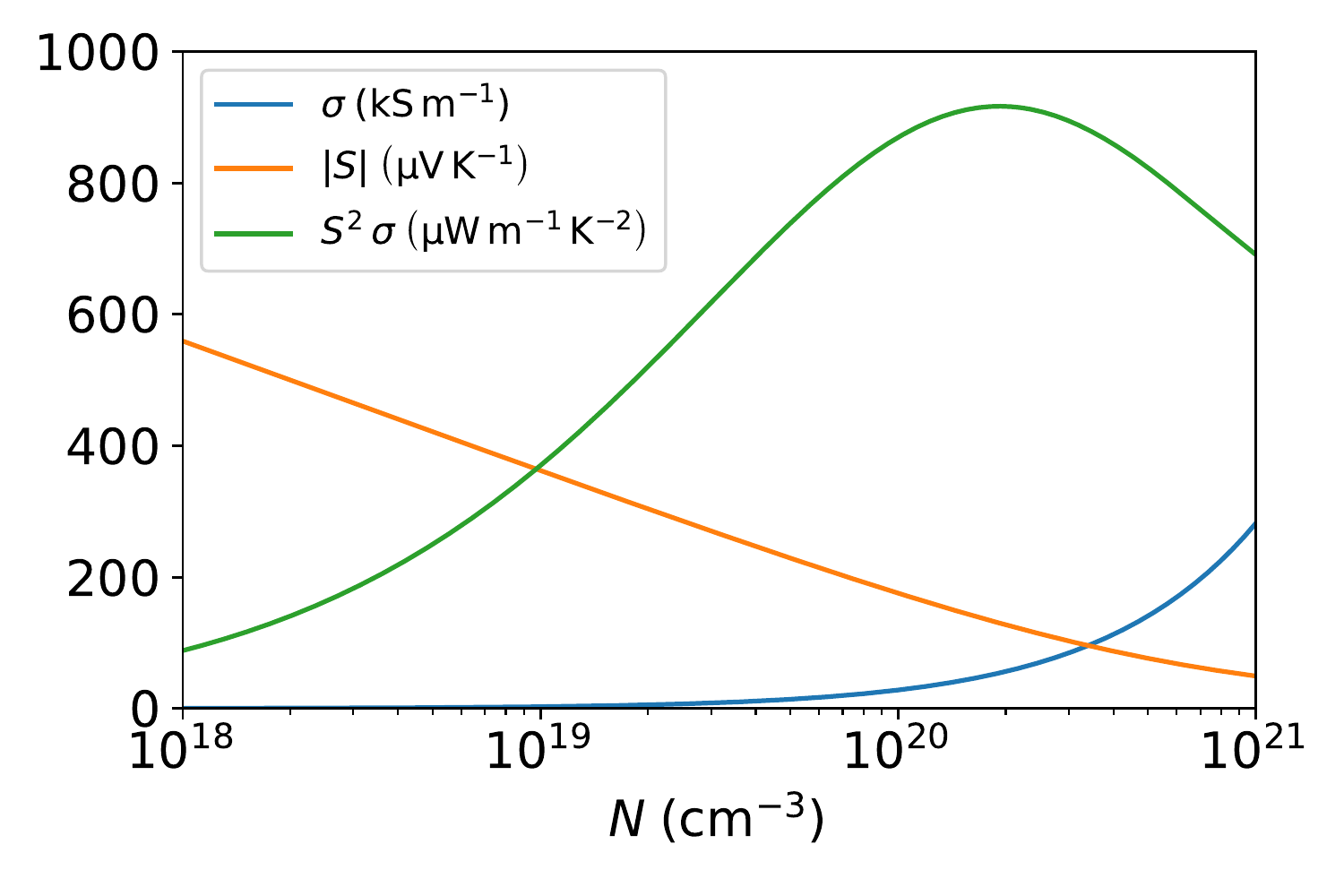}
\end{center}
\caption{$S$, $\sigma$, and the thermoelectric $PF$ as a function of carrier concentration. The transport coefficients have been evaluated using a parabolic band model with $m^*=m_e$. The temperature and relaxation time were set to $T=500$~K and $\tau=10^{-14}$~s, respectively. }
\label{fig:3DHEG2}
\end{figure}

The simplest way to illustrate the library functionality of \texttt{BoltzTraP2} is the parabolic band model. Consider a dispersion relation expressed as, 
\beq
\varepsilon(k)=\frac{\hbar^2k^2}{2m^*}
\label{eq:model}
\eeq
where $m^*$ is the effective mass of the electrons in the band. For an isotropic  parabolic band model we can replace the outer
product of group velocities, Eq.~\eqref{eq:sigma}, by $k/m^*$
 and the volume element $\id\bk$ in the three dimensional volume integral in Eq.~\eqref{eq:sigma} by $4\pi k^2\id k$, thereby obtaining analytic expressions for $n(\varepsilon)$ and $\sigma(\varepsilon)$ 
\begin{gather}
n(\varepsilon)=\frac{1}{4\pi^2}\biggl(\frac{2m^*}{\hbar^2}\biggr)^{3/2} \varepsilon^{1/2}  \\
\sigma(\varepsilon)= \frac{1}{3\pi^2}\frac{\sqrt{2m^*}}{\hbar^3} \tau  \varepsilon^{3/2}
\label{eq:coeff2}
\end{gather}
Evaluating the carrier concentration, Eq.~\eqref{eq:carrier}, and the transport coefficients, Eqs.~\eqref{eq:coeff_E}-\eqref{eq:Seebeck}, leads directly to the  the famous plot Fig.~\ref{fig:3DHEG2}. For comparison we have created a single parabolic band numerically on a $25\times 25\times 25$ \bk-mesh for a cubic unit cell with $a=5$~\AA. The band was interpolated onto a mesh containing 5 times the points. The resulting transport coefficients are indistinguishable from the analytic in Fig.~\ref{fig:3DHEG2}.

\subsection{Inclusion of momentum matrix elements. Silicon.}
The band structure of silicon makes one of the simplest examples that is not trivial for an interpolation scheme. The CBM of Si is made up of pockets found along the six-fold degenerate $\Gamma-X$ line. Furthermore, in the standard FCC setting it has a non-symmorphic space group
so that the bands can cross at the zone boundary. A crossing at the zone boundary will mean that the bands will not necessarily touch the zone-boundary ``horizontally''. A purely Fourier-based interpolation scheme, as the one used in \texttt{BoltzTraP2} can give false derivatives at these points, meaning that a very fine $k$-mesh can be necessary to obtain converged results.

\begin{figure}[htb]
\begin{center}
\includegraphics[width=.9\columnwidth]{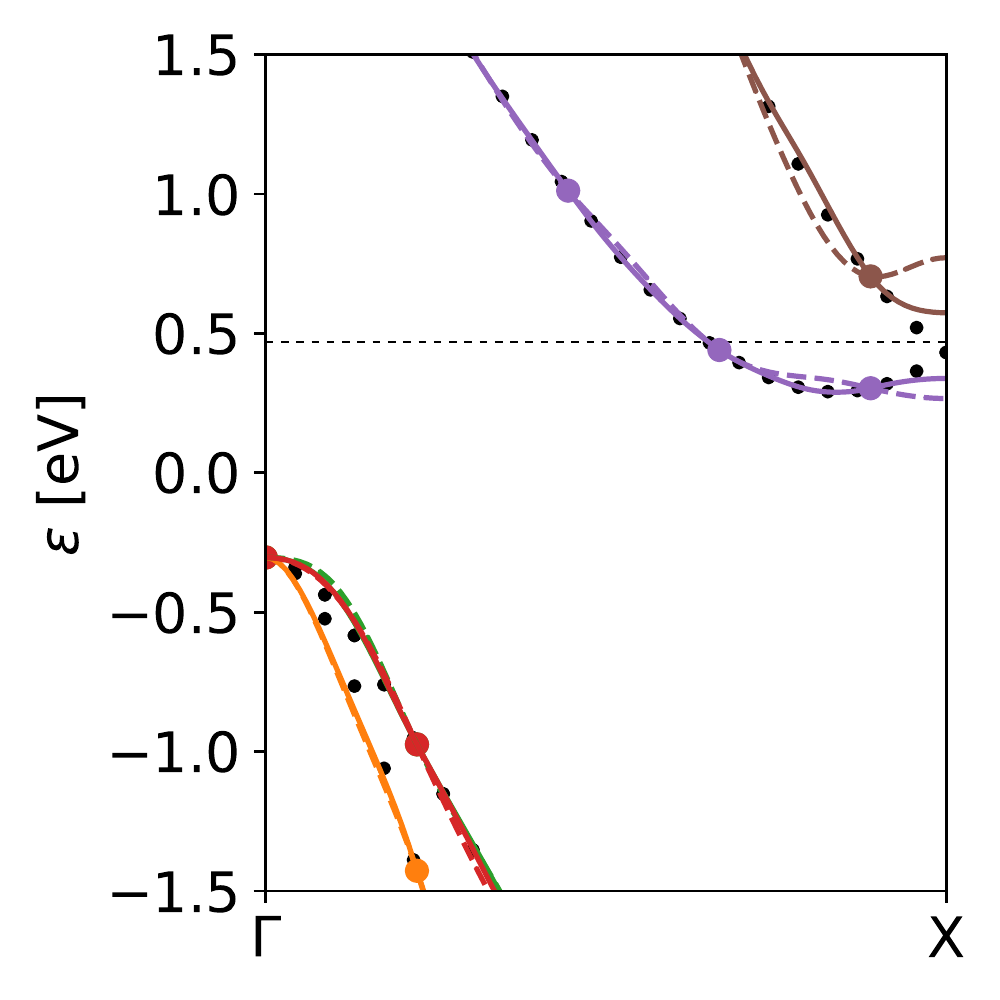}
\end{center}
\caption{Silicon band edges along the $\Gamma-X$ line. The black points are calculated points along this specific direction. The colored lines correspond to the interpolated bands based on a coarse $9\times9\times9$ $k$-point mesh. The points belonging to this mesh are marked with larger colored points. The full lines are obtained by including the momentum matrix elements in the fit and the dashed use only the eigenvalues. Thin dashed line: chemical potential used below in Fig.~\ref{fig:Si_conv}}
\label{fig:Si_pocket}
\end{figure}

The CBM pocket found along the $\Gamma-X$ line, which will dominate the transport properties of $n$-doped Si, is illustrated in Fig.~\ref{fig:Si_pocket}. Fig.~\ref{fig:Si_pocket} compares the result of a usual DFT calculation of a band structure, with a fine set of $k$-points along a specific direction, with that obtained by the analytic interpolation of a coarse $9\times9\times9$ $k$-point mesh, Eq.~\eqref{eq:Fourier}. A  $9\times9\times9$-mesh corresponds to only 35 $k$-points in the irreducible part of the Brillouin zone (IBZ). It is obviously not a fine $k$-mesh, that would typically be calculated non-self-consistently for transport calculations\cite{Madsen_JACS06}, and corresponds to a typical $k$-mesh used for a self-consistent DFT calculation. Furthermore, as the lowest conduction bands are degenerate at the $X$-point, Fig.~\ref{fig:Si_pocket}, the non-symmorphic space group does result in the derivatives of interpolated bands being incorrect at this point. However, Fig.~\ref{fig:Si_pocket} also illustrates how the modified Lagrangian, Eq.~\eqref{eq:Lagrangian}, forces the fit to reproduce the exact derivatives at the calculated points. Thereby, both the position and derivatives of the pocket are well reproduced. On the other hand, if only the eigenvalues are included in the fit, this mesh is obviously too coarse and fails to reproduce either the position or the derivatives at the pocket (purple dashed line in Fig.~\ref{fig:Si_pocket}).

\begin{figure}[htb]
\begin{center}
\includegraphics[width=.9\columnwidth]{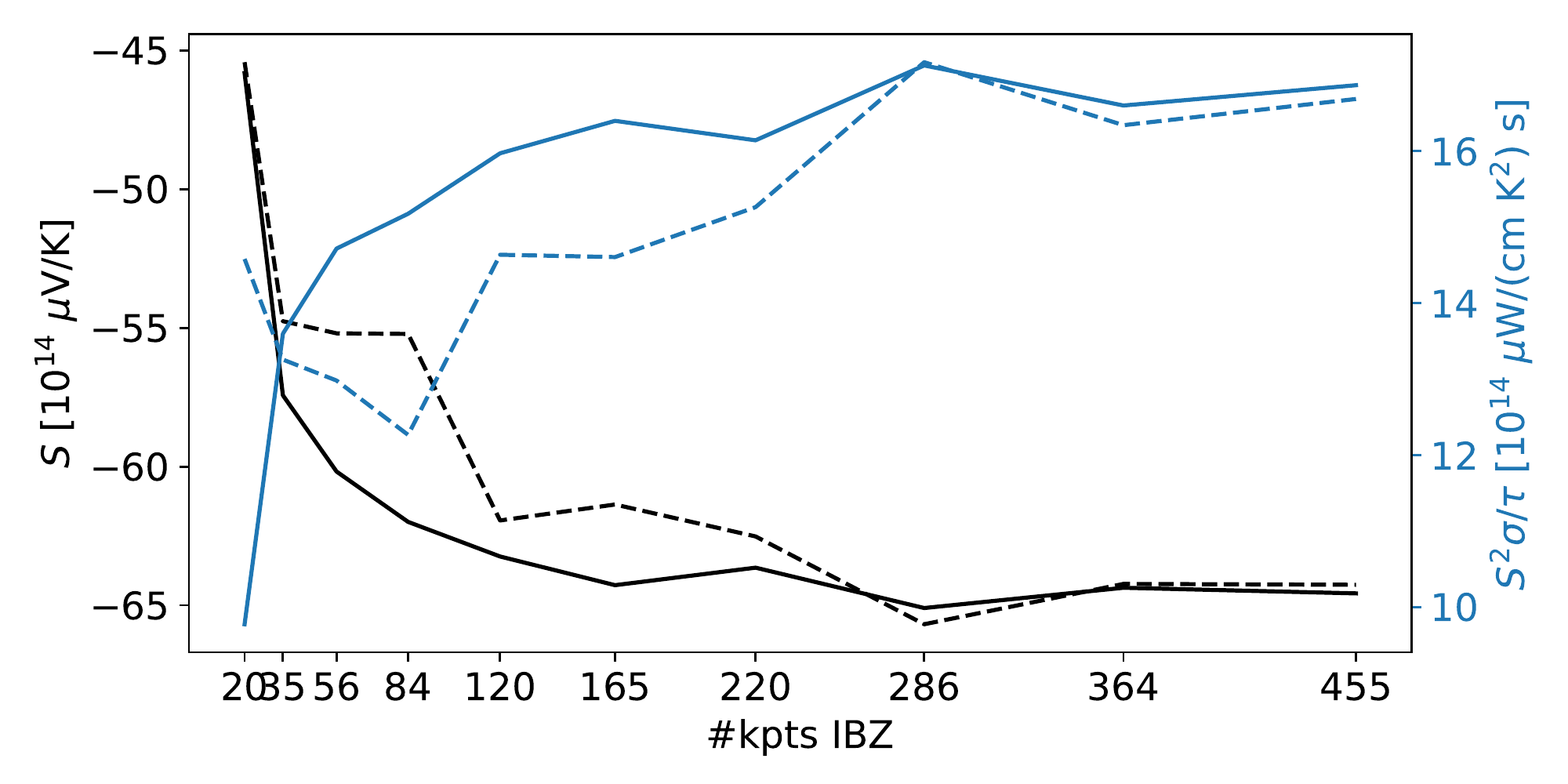}
\end{center}
\caption{Convergence of the Seebeck coefficient and thermoelectric power factor as a function of number $k$-points in the irreducible Brillouin zone. The full lines are obtained by including the momentum matrix elements in the fit and the dashed use only the eigenvalues.}
\label{fig:Si_conv}
\end{figure}

The impression obtained graphically in Fig.~\ref{fig:Si_pocket} is quantified in Fig.~\ref{fig:Si_conv}. The Seebeck coefficient and the thermoelectric power factor, $S^2\sigma/\tau$, are calculated at a chemical potential close to the CBM (marked by the dashed line in Fig.~\ref{fig:Si_pocket}) using the CRTA, Eqs.~\eqref{eq:sigma}-\eqref{eq:Seebeck}. It is seen how the results obtained by the modified Lagragian show both a faster and more systematic trend, reaching convergence at about half the number of $k$-points needed when the derivatives are not included in the fit.

\subsection{State dependent relaxation time in Lithium.}
\begin{figure}
\begin{center}
\includegraphics[width=.9\columnwidth]{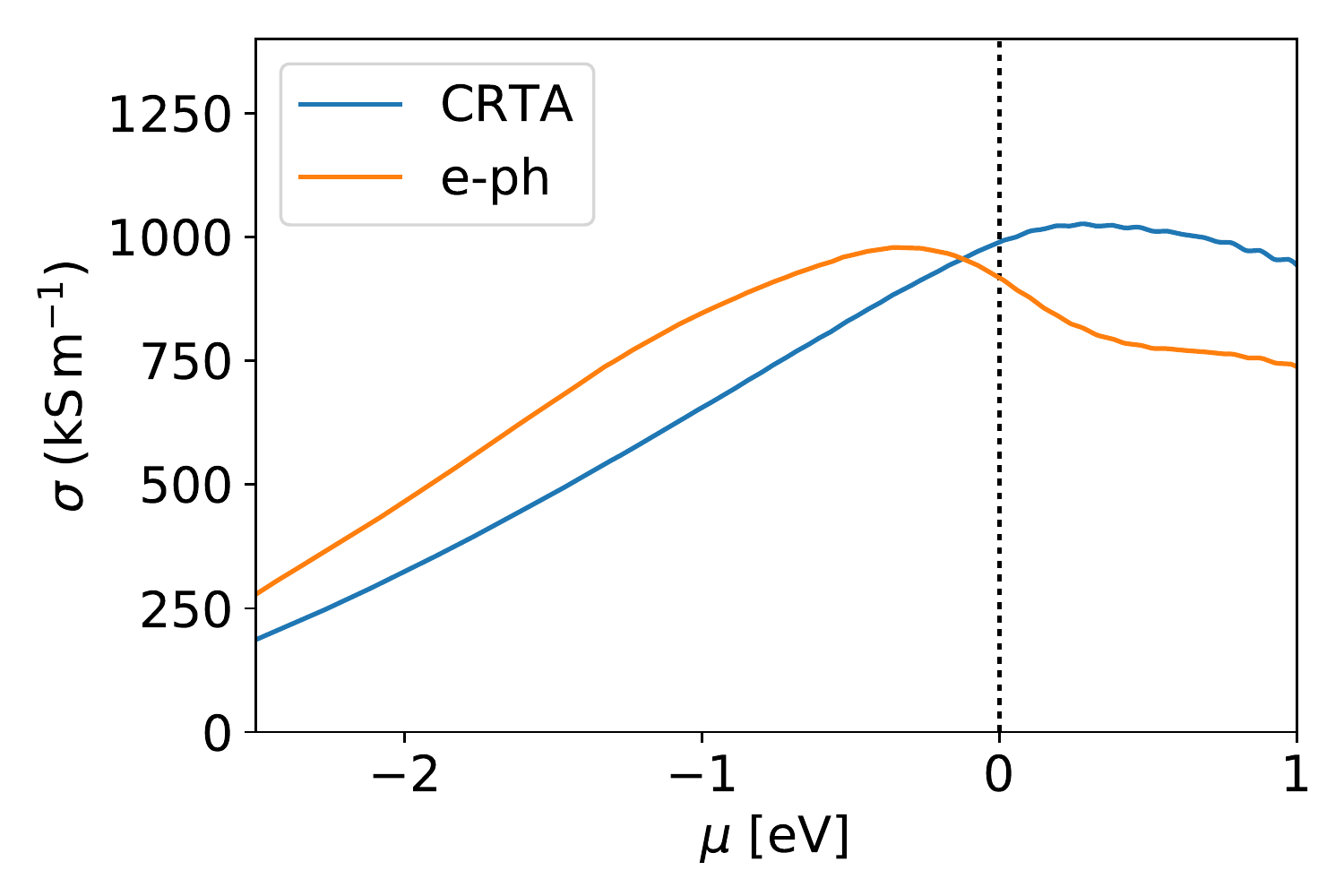}
\end{center}
\caption{Calculated conductivity of bcc-Li using the CRTA and band and momentum dependent relaxation times due to electron-phonon coupling. The constant relaxation time was set to the average of the calculated $\tau^{ep}_{n\bk}$ ($\tau = 1.05\cdot 10^{-15}$~s)}
\label{fig:Li}
\end{figure}

In \texttt{BoltzTraP2} the interpolation and integration steps are more explicitly decoupled than in \texttt{BoltzTraP}. This allows the interpolation capabilities of the code to be used to match quantities represented on different grids. We illustrate this possibility by considering the transport distribution function of bcc-Lithium. As a by-product of the calculations in Xu et al\cite{Xu_PRL14}, we can obtain the computationally costly relaxation times due to electron-phonon scattering, $\tau^{ep}_{n\bk}$, on a relatively coarse $24\times24\times24$ $\bk$-point mesh. The quasi-particle energies were interpolated onto a grid containing 60 times the number of $\bk$-points. Fig.~\ref{fig:Li} illustrates how this leads to a positive slope of the CRTA transport distribution function at the Fermi level. Consequently, Eqs.~\eqref{eq:coeff_E} and \eqref{eq:Seebeck}, we find a negative Seebeck coefficient of $S=-2\ \mu$V/K at 300~K as in Ref.~\cite{Xu_PRL14}. Using our interpolation scheme, the calculated band- and momentum-dependent relaxation times were interpolated onto the same fine grid. The inclusion $\tilde{\tau}^{ep}_{n\bk}$ leads to a change of slope at the Fermi level and consequently a positive Seebeck coefficient ($S=+4\ \mu$V/K at 300~K vs $+13\ \mu$V/K within the Lowest Order Variational Approximation in \cite{Xu_PRL14}).
 
The obtained CRTA transport distribution function, Fig.~\ref{fig:Li}, is in good agreement with a more ``usual'' \texttt{BoltzTraP} type calculation where energies calculated on $58\times 58\times58$-grid were interpolated onto a grid containing four times as many $\bk$-points. This lends credibility to the interpolation of the scattering rates as well. 

The interpolated quasi-particle energies are usually considered to be independent from parameters such as temperature and Fermi level, and hence the interpolation does not need to be repeated (only the integration), for instance, to estimate thermoelectric coefficients for a different doping level or temperatures. The direct interface to the interpolation routines make it straightforward to interpolate the quasi-particle energies once and for all, and to avoid duplication of work when interpolating a temperature dependent $\tau$.

\section{Conclusion}
We have presented a new a software package, \texttt{BoltzTraP2}, based mainly on Python 3. The methodology is based on a smoothed Fourier expression for periodic functions and uses only the band and $k$-dependent quasi-particle energies as well as the intra-band optical matrix elements and scattering rates as input. The Onsager transport coefficients have been evaluated for a simple periodic band, as well as Silicon and Lithium using the linearized Boltzmann transport equation. The code can be used via a command-line interface and as a Python module.

\section{Acknowledgements}
MJV acknowledges support from the Communaut\'e fran\c{c}aise de Belgique through an ARC grant (AIMED 15/19-09) and the Belgian Fonds National de la Recherche Scientifique FNRS, under grant number PDR T.1077.15-1/7. GKHM and JC acknowledge support from M-era.net through the ICETS project (DFG: MA 5487/4-1) and the EU Horizon 2020 grant No. 645776 (ALMA).
Computational resources have been provided by the Consortium des Equipements de Calcul Intensif en F\'{e}d\'{e}ration Wallonie Bruxelles (CECI), funded by FRS-FNRS G.A. 2.5020.11; the Tier-1 supercomputer of the F\'{e}d\'{e}ration Wallonie-Bruxelles, funded by the Walloon Region under G.A. 1117545; the PRACE-3IP DECI grants, on ARCHER and Salomon (ThermoSpin, ACEID, OPTOGEN, and INTERPHON 3IP G.A. RI-312763) and the Vienna Scientific Cluster (project number 70958: ALMA). 
%


\end{document}